\documentclass[12pt]{article}
\usepackage{epsfig}

\usepackage{amsmath}
\usepackage{verbatim}
\usepackage{graphicx}
\usepackage{caption}
\usepackage{subcaption}

\newcommand{\gev}{{\unskip\,\text{GeV}}}
\newcommand{\tev}{{\unskip\,\text{TeV}}}

\newcommand{\eq}[1]{Eq.~(\ref{#1})}
\newcommand{\bib}[1]{Ref.~\cite{#1}}

\newcommand{\fig}[1]{Fig.~\ref{#1}}
\newcommand{\tab}[1]{Table~\ref{#1}}
\newcommand{\sect}[1]{Section~\ref{#1}}

\newcommand{\bea}{\begin{eqnarray}}
\newcommand{\eea}{\end{eqnarray}}
\newcommand{\bal}{\begin{aligned}}
\newcommand{\eal}{\end{aligned}}

\newcommand{\crn}{\nonumber \\}

\newcommand{\fr}{\frac}

\def\Title#1{\begin{center} {\Large {\bf #1} } \end{center}}
\def\Abstract#1{\noindent {\small { #1} } }

\begin{document}
\rightline{HU-EP-16/41}
\rightline{IFIRSE-TH-2016-1}

\Title{A theoretical status of weak gauge boson pair production at the LHC\footnotemark[1]}
\footnotetext[1]{Talk presented at the XXXVI international symposium on Physics in Collision, 
September 2016, Quy Nhon, Vietnam.}

\bigskip

\begin{raggedright}  

{\it LE Duc Ninh\index{LE Duc Ninh}\\
Humboldt-Universit\"{a}t zu Berlin\\
Newtonstrasse 15, D-12489 Berlin, Germany\\
and\\ 
Institute For Interdisciplinary Research in Science and Education (IFIRSE)\\ 
ICISE \\
Ghenh Rang, 590000 Quy Nhon, Vietnam\\
\smallskip
Email: ldninh@ifirse.icise.vn}
\bigskip
\end{raggedright}

\Abstract{{\bf Abstract:} Weak gauge boson pair production is an important process at the LHC 
  because it probes the non-Abelian structure
  of electroweak interactions and it is a background process for many new physics 
  searches, and with enough statistics we can perform comparisons 
  between measurements and theoretical calculations for different but correlated 
  observables. In these proceedings, we present a theoretical status including state-of-the-art results from recent calculations 
  of higher-order QCD and electroweak corrections.}

\section{Introduction}
Since the discovery of the Higgs boson in 2012 by ATLAS \cite{Aad:2012tfa} and CMS \cite{Chatrchyan:2012xdj}, there has been no clear evidence of new physics at the LHC. 
This naturally leads to the focus on precision physics. 
If done carefully and correctly, precision physics will help 
us to identify new physics beyond the Standard Model (SM). 
We can estimate how high precision can be achieved, but unfortunately we 
do not know how strong are new physics effects. 

LHC at $13\tev$ brings great opportunities. It opens up new channels 
which have never been explored so far. These include top quarks, gauge bosons and 
Higgs bosons in the final state. In the following, we will discuss the case of weak 
gauge boson pair production, i.e. $W^+W^-$, $W^\pm Z$ and $ZZ$, in the SM.
  
\begin{figure}[htb]
\begin{center}
\begin{minipage}[c]{6.0cm}
\epsfig{file=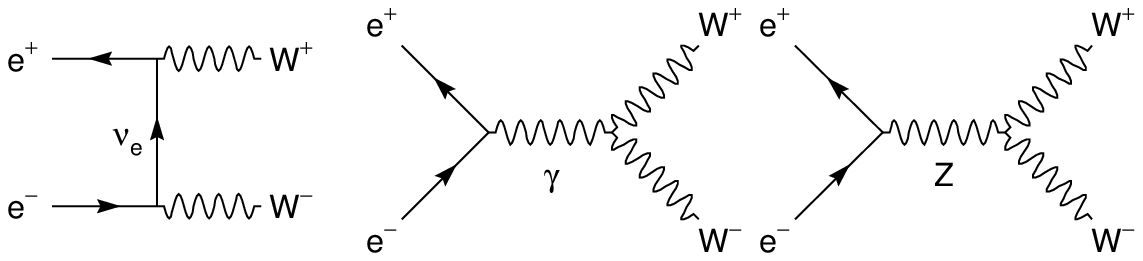,width=6.0cm}
\end{minipage}
\begin{minipage}[c]{8.0cm}
\epsfig{file=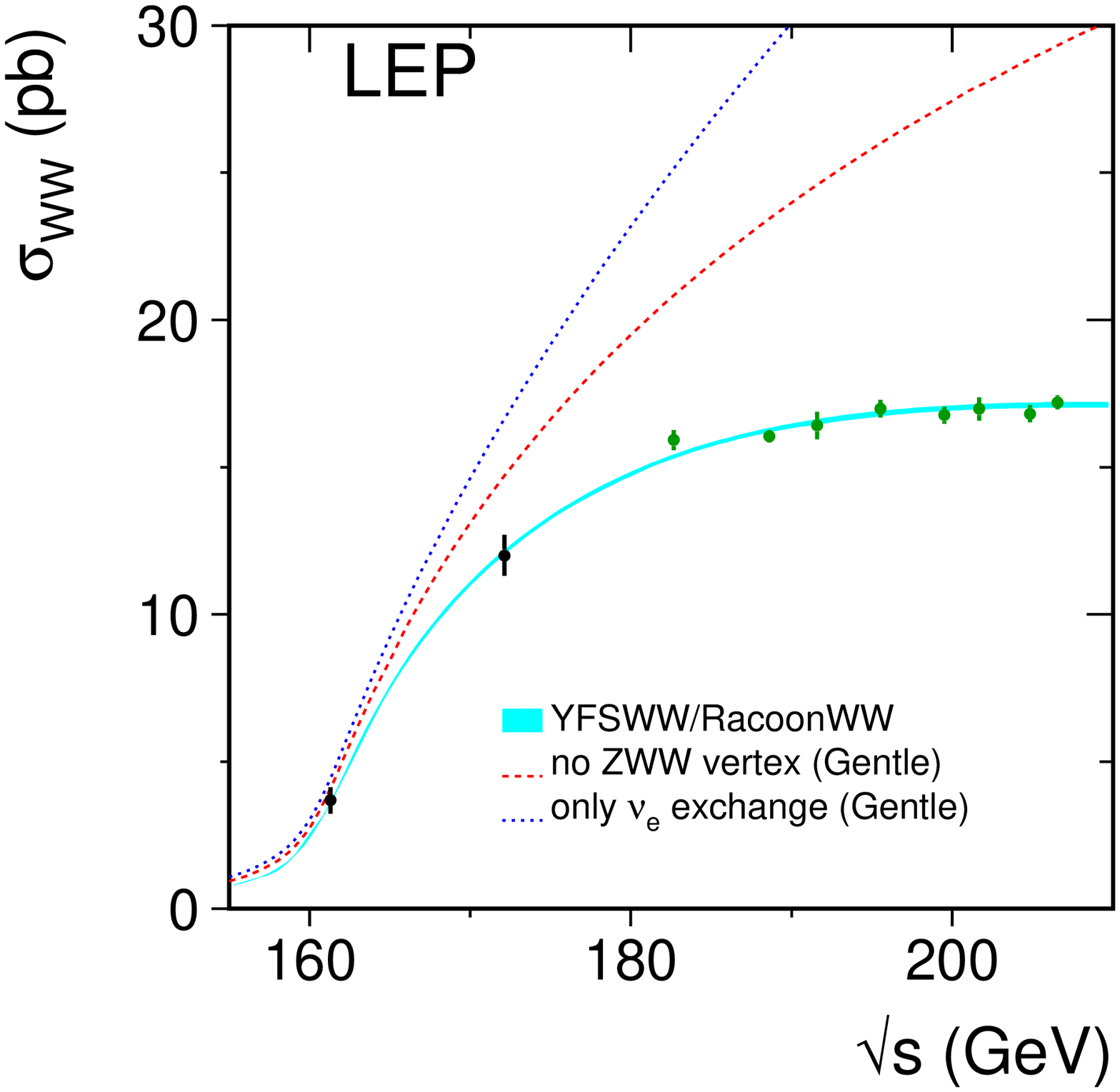,height=7.0cm}
\end{minipage}
\caption{Left: Born-level Feynman diagrams. 
Right: total cross section in pb as a function of center-of-mass energy at LEP2.  
Plots taken from \bib{LEP-2}.}
\label{fig:LEP2_WW}
\end{center}
\end{figure}

Weak gauge boson pair production mechanisms
provide an important test of the non-Abelian gauge structure of the SM. 
It is a good way to check the trilinear gauge couplings among the $W$, $\gamma$ and
$Z$ bosons. The key point here is gauge invariance. If this is broken, e.g. the s-channel 
diagram is removed, then unitarity will be violated and agreement with measurements will be destroyed, 
as seen from \fig{fig:LEP2_WW} at LEP2 \cite{LEP-2}. 
One interesting question arises in this context. Asumming that gauge invariance is preserved, 
to what extend we can modify the SM gauge couplings such that agreement with 
experimental data can still be achieved?    

In addition, diboson production are important backgrounds to many studies 
such as Higgs boson production. It is therefore mandatory to have a 
rigorous theoretical understanding and precise calculations. 

The QCD next-to-leading order (NLO) corrections to massive gauge boson pair
production have been known for a long time~\cite{Mele:1990bq, Ohnemus:1990za, Ohnemus:1991kk, WZ-QCD-NLO-1,WZ-QCD-NLO-2,WW-QCD-NLO}. 
Recently, next-to-next-to-leading order (NNLO) QCD calculations have been completed 
for $ZZ$~\cite{Cascioli:2014yka}, $WW$~\cite{Gehrmann:2014fva,Grazzini:2016ctr} and $WZ$~\cite{Grazzini:2016swo}.  
Concerning EW corrections, results at the on-shell level were obtained in \cite{WW-EW-Kuhn,Bierweiler:2013dja,Baglio:2013toa}, 
and recently full results including off-shell effects and spin correlations for leptonic final state 
have been calculated in \cite{Biedermann:2016yvs,Biedermann:2016guo} for the case of 
electrically neutral final state. 
It is now a good time to review our current understanding of massive gauge boson pair production at the LHC up to the level of NNLO QCD and NLO EW, with or without leptonic decays. The important topic of anomalous couplings is not discussed in this work.

\section{Born level and generalities}
The process
\bea 
pp \to VV' + X,
\label{eq:proc_VV_OS}
\eea
where $V,V' = W,Z$ has two important features at Born level, with two quarks in the initial state.  
The amplitude is of purely electroweak nature, being independent of $\alpha_s$. 
The amplitude is strongly constrained by gauge invariance. Notably, the vanishing of 
the $ZZZ$ coupling at tree level and only $WWZ$ and $WW\gamma$ are allowed at tree level in the SM. 
At the LHC, we can classify \eq{eq:proc_VV_OS} into two classes according to the total 
electrical charge of the final state: $WW$ and $ZZ$ have $Q = 0$, while $W^\pm Z$ have $Q = \pm 1$. 

That is a simplified picture at the on-shell level. In reality, the mean lifetime of $W$ bosons is about $3.2\times 10^{-25}$s 
and $Z$ about $2.6\times 10^{-25}$s \cite{Olive:2016xmw}. The branching ratio of $W$ decay into charged leptons (muon or electron) is 
about $21.3\%$ and $Z$ about $6.7\%$ \cite{Olive:2016xmw}. It is therefore possible to use the fully leptonic decay modes to do 
precision physics at the LHC. The fully hadronic decay modes, even though having larger branching ratios, suffer from large QCD backgrounds 
and the issue that hadronic mass resolution is limited. 
The semileptonic chanels can however be interesting and have been investigated at ATLAS \cite{Aaboud:2016uuk} and CMS \cite{CMS:2016djf}. 
When $V$ is produced with large transverse momentum, its hadronic decay products are highly collimated, leading to a single 
jet. A recent ATLAS study \cite{Aad:2015rpa} shows that using jet substructure techniques 
can provide a $50\%$ efficiency for identifying $W$ bosons with $p_T > 200\gev$. 

From leptons in the final states, the so-called $ZZ$, $WW$, and $WZ$ signals are then 
defined using cuts to select leptons which likely originate from the $Z$ or $W$ bosons. 
To compare with measurements we must do the same from theory side. On-shell approximations 
are useful to understand the structure of radiative corrections, but are not good enough for 
precise comparisons with data. 

\section{NLO and NNLO QCD corrections}
In this section, we mainly try to understand the structure of QCD radiative corrections. 
For this purpose, the on-shell $VV'$ final state are considered.

\label{sec:NLO_NNLO_QCD}
\begin{figure}[htb]
\begin{center}
\begin{minipage}[c]{6.0cm}
\epsfig{file=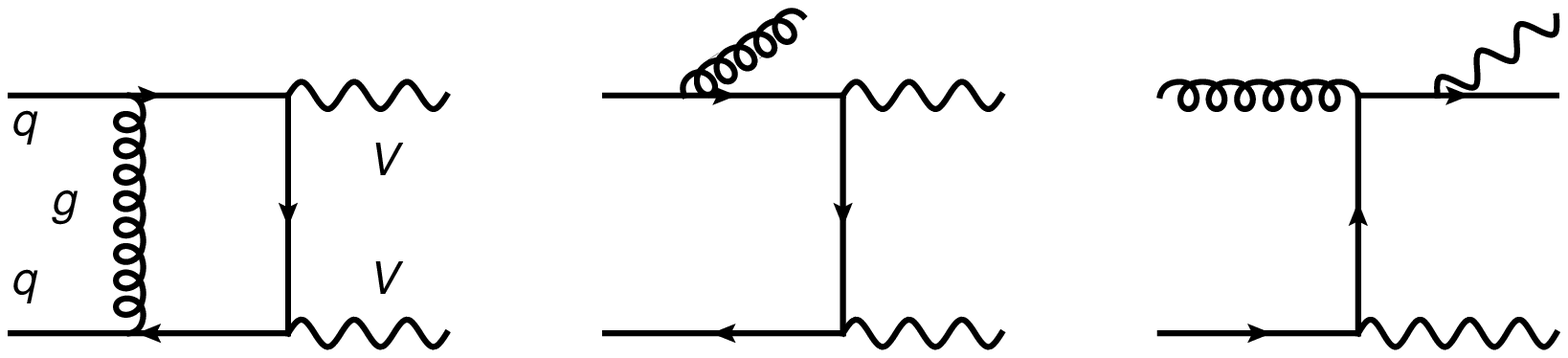,width=6.0cm}
\end{minipage}
\begin{minipage}[c]{8.0cm}
\epsfig{file=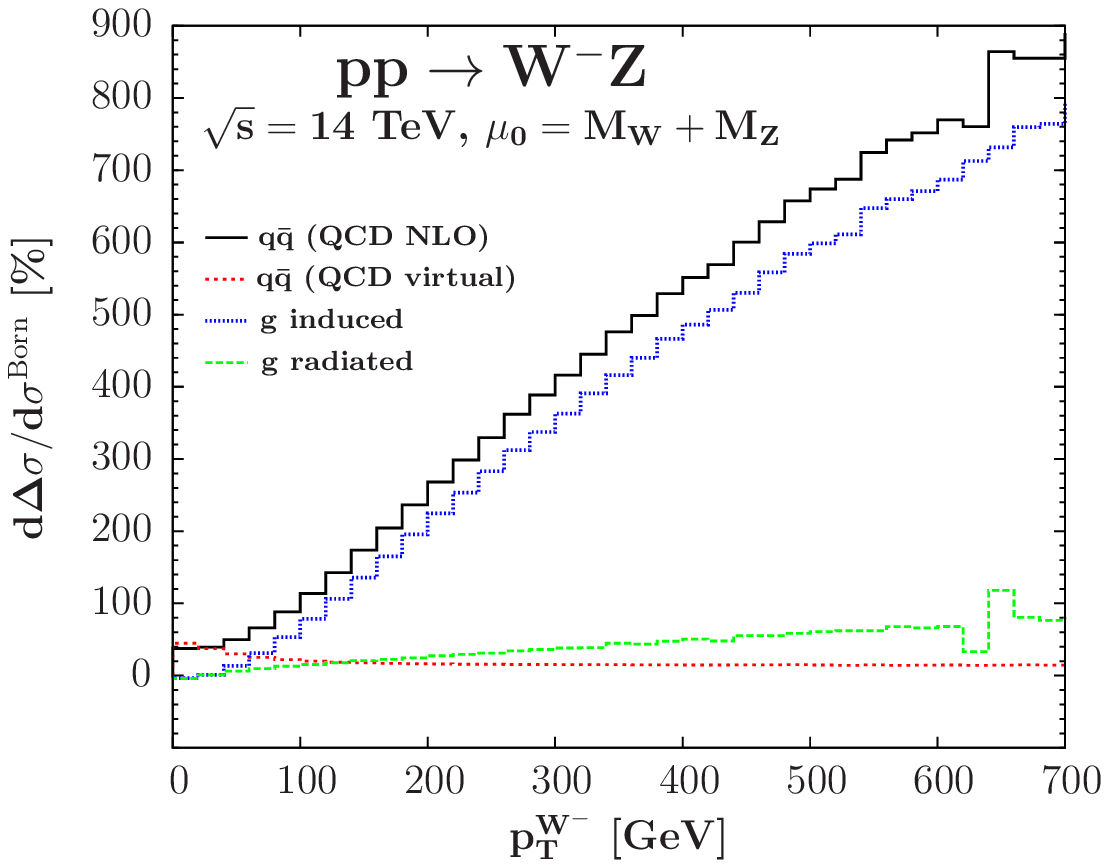,height=6.0cm}
\end{minipage}
\caption{Left panel: Representative Feynman diagrams of NLO QCD corrections. 
The left diagram corresponds to virtual correction, the middle to real-gluon emission, 
and the right to quark-gluon induced contribution. 
Right panel: the virtual, gluon radiated, quark-gluon induced contributions to the $p_T^{W^-}$ distribution, 
normalized to the Born result, for the process $pp\to W^- Z$. Their sum (labeled QCD NLO) is also shown.
Right plot taken from \bib{Baglio:2013toa}, where precise definition of the individual corrections are given.}
\label{fig:diag_NLO_QCD_with_pT_WZ}
\end{center}
\end{figure}
The NLO QCD contribution of order $\mathcal{O}(\alpha_s)$ 
takes into account virtual and real emission corrections with one additional
parton in the final state. Representative Feynman diagrams are shown in \fig{fig:diag_NLO_QCD_with_pT_WZ} (left). 
The important point to notice is that two new parameters enter the game: $\alpha_s$ and 
gluon parton distribution function (PDF). NLO corrections change not only the kinematics picture but also the list of 
subprocesses. As can be seen in \fig{fig:diag_NLO_QCD_with_pT_WZ} (right), for the case of $W^- Z$ production, the quark-gluon induced processes 
provides a dominant contribution at large $p_T$. 
This contribution is huge, about $8$ times larger than the Born contribution at $p_T = 700\gev$. 
This is a proof that QCD radiation can introduce large corrections, not only for kinematical distributions but also 
for total cross section (see \tab{tab:total_xsection}). 
To understand this, it is important to note that the LO picture consists of 
only $q\bar{q}$ subprocesses. 
If we look into the quark PDFs, we see also gluon PDF and $\alpha_s$ there. 
But remember that these parameters are confined there in the low-energy region, where only soft and collinear splittings are taken into account in the evolution of PDFs. 
At large $p_T$, $\alpha_s$ is small due to asymptotic freedom and $g(x)$ is small due to large $x$ 
(which is proportional to $p_T/\sqrt{s}$, with $\sqrt{s}$ being the center-of-mass energy of the proton-proton collision). 
So, why do we get larger correction with increasing $p_T$?

The answer was partially known a long time ago \cite{WZ-QCD-NLO-2, WW-QCD-NLO}.
The explanation, for the case of $qg \to ZW q$, was due to the following mechanism. 
First a hard $Z$ is produced, recoiling against the quark, and then the final quark 
radiates a soft $W$. The radiation of a soft $W$ gives rise to a correction factor of 
the form $\alpha\log^2(p_{T,Z}^2/M_W^2)$, which is responsible for the behavior shown in 
\fig{fig:diag_NLO_QCD_with_pT_WZ} (right). 

\begin{figure}[htb]
\begin{center}
\begin{minipage}[c]{6.5cm}
\epsfig{file=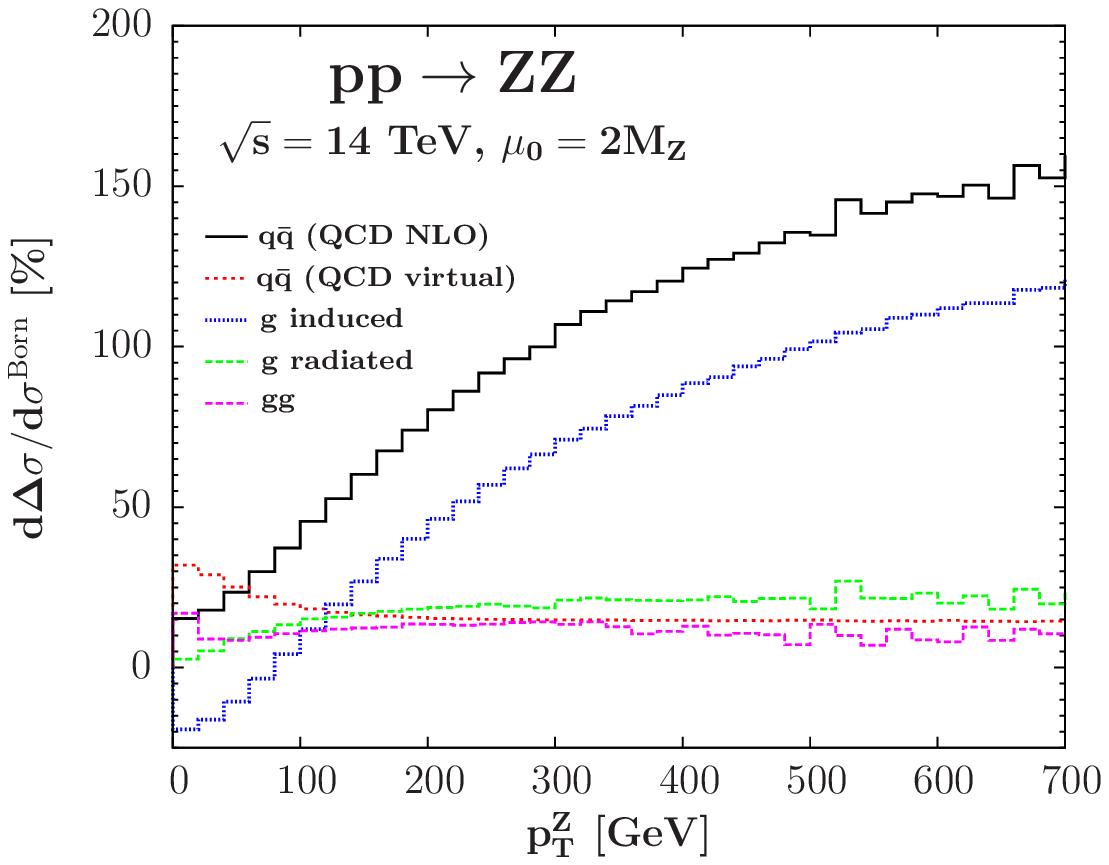,width=6.1cm}
\end{minipage}
\begin{minipage}[c]{6.0cm}
\epsfig{file=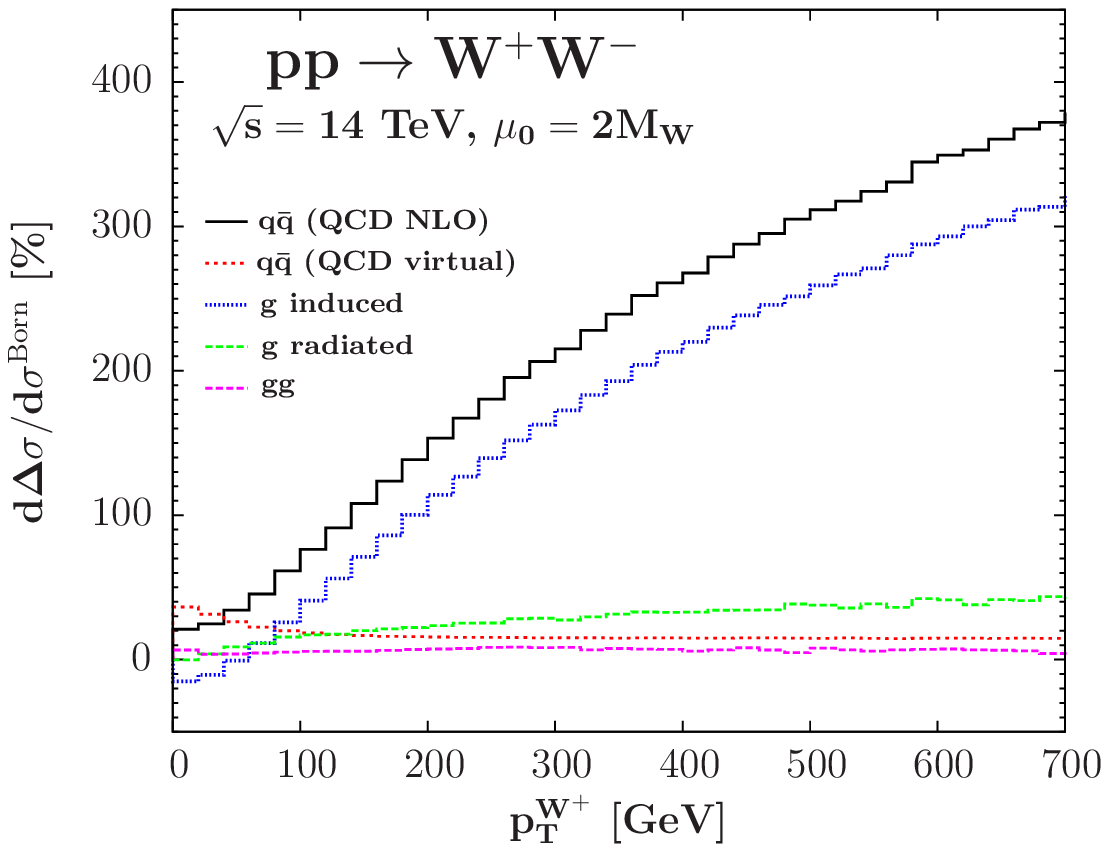,width=6.0cm}
\end{minipage}
\caption{Similar to \fig{fig:diag_NLO_QCD_with_pT_WZ} but for $ZZ$ (left) and 
$WW$ (right) cases. Plots taken from \bib{Baglio:2013toa}.}
\label{fig:dist_pT_NLO_QCD_ZZ_WW}
\end{center}
\end{figure}

So far so good. Let's transfer the above understanding to the case of $ZZ$ and $WW$. 
The results are shown in \fig{fig:dist_pT_NLO_QCD_ZZ_WW}. 
We can see that, qualitatively, the pictures look the same. The dominant 
correction at large $p_T$ is the $qg$ channel and it has the same behavior 
as above understood. However, if we look carefully at the magnitude of the corrections, 
we notice that it is much larger for $WW$ than for $ZZ$, and largest for $WZ$. 
The difference between $M_W$ and $M_Z$ is not enough to explain this, as 
observed in \cite{QCD-NLO-exclusive}. So, something must be missing.

The missing piece is something obvious. 
The soft $W$, in the $WZ$ case, can 
be radiated not only from the final quark (as above said) but also from the $Z$ boson and the initial-state quark. 
All possibilities must be taken into account to have a gauge invariant result. 
This helps us to see that the $ZZ$ case is special as there are only two possibilities of $Z$ radiation, 
from initial or final quarks. For the case of $WW$ and $WZ$, one gauge boson can be radiated additionally from the other gauge boson. 
A detailed calculation using the leading logarithmic approximation at large $p_T$ was done in \bib{Baglio:2013toa}. 
The results read
\begin{align}
d\sigma^{qg \to ZZ q}
&= c^q_{ZZ} d\sigma^{qg \to Z q}_L
\fr{\alpha}{2\pi}\log^2\left[\fr{(p_T^{Z})^2}{M_Z^2}\right],\label{eq_log2_zz}\\
d\sigma^{qg \to W^+W^- q}
&= c^q_{WW}d\sigma_L^{qg \to Z q}
\fr{\alpha}{2\pi}\log^2\left[\fr{(p_T^{W^+})^2}{M_W^2}\right],\label{eq_log2_ww_pTWpm}\\
d\sigma^{ug \to W^+Z d}
&= c^u_{WZ} d\sigma_L^{ug \to Z u}
\fr{\alpha}{2\pi}\log^2\left[\fr{(p_T^{W^+})^2}{M_Z^2}\right],\label{eq_log2_wpz}\\
d\sigma^{dg \to W^-Z u}
&= c^d_{WZ} d\sigma_L^{dg \to Z d}
\fr{\alpha}{2\pi}\log^2\left[\fr{(p_T^{W^-})^2}{M_Z^2}\right],\label{eq_log2_wmz}
\end{align}
where the subscript $L$ means that all quarks are left-handed and the coefficients $c_{VV'}^q$ 
depend on the quark charges ($Q_q$ and hypercharge $Y_q$). Numerically, we get
\begin{align}
c^u_{ZZ} &\approx 0.18,\quad c^d_{ZZ}\,\, \approx 0.26, \crn
c^u_{WW} &\approx 3.53,\quad c^d_{WW} \approx 2.40, \crn
c^u_{WZ} &\approx 4.13,\quad c^d_{WZ}\, \approx 2.81. \label{eq:c_VV_values}
\end{align}
We can now see better why the quark-gluon induced correction is different for the three 
processes. It is noted that Eqs.~(\ref{eq_log2_wmz}), (\ref{eq_log2_zz}), (\ref{eq_log2_ww_pTWpm}) are only for 
the numerators of the ratios shown in Figs.~\ref{fig:diag_NLO_QCD_with_pT_WZ} and 
\ref{fig:dist_pT_NLO_QCD_ZZ_WW}. To fully explain these plots we have to compare also 
the denominators. By doing so we recognize that PDFs also play a role. 

The above line of reasoning helps us to notice that the NLO picture is still 
not complete. The gluon-gluon fusion mechanism is not there. This may be important 
due to large gluon PDF. Moreover, the process $gg \to \bar{q}q' VV'$ occurring at NNLO QCD, 
introduces a new mechanism, where both gauge bosons are softly radiated off the final quarks. 
This brings large double-double logarithmic effects when the two quarks are produced with high $p_T$, 
recoiling against each other. However, the amplitude of $gg \to \bar{q}q'$ is smaller at increasing $p_T$. 
So we are seeing here two competing effects. The question is: how large is NNLO QCD correction?

To answer that question, we first look at the corrections to the total cross section. 
This does not help us to understand what happens at large $p_T$, because the main contribution 
to the total cross section comes from low $p_T$ region. Nevertheless, it is also important 
to examine the small $p_T$ region and this simple observable can help us to understand the QCD corrections better. 
The total cross sections at LO, NLO QCD, and NNLO QCD for on-shell production at $13\tev$ LHC are given 
in \tab{tab:total_xsection}.
\begin{table}[th]
\begin{center}
\begin{tabular}{l|ccccc}  
 &  LO & NLO QCD & NNLO QCD & NLO/LO & NNLO/NLO \\ 
\hline
 $ZZ$  &      $9.9^{\footnotesize +4.9\%}_{\footnotesize -6.1\%}$     &     $14.5^{\footnotesize +3.0\%}_{\footnotesize -2.4\%}$      &  $16.9^{\footnotesize +3.2\%}_{\footnotesize -2.4\%}$   &   $1.5$   &   $1.2$  \\
\hline
 $W^+ W^-$ &  $67.2^{\footnotesize +5.5\%}_{\footnotesize -6.7\%}$    &     $106.0^{\footnotesize +4.1\%}_{\footnotesize -3.2\%}$     &  $118.7^{\footnotesize +2.5\%}_{\footnotesize -2.2\%}$  &   $1.6$   &   $1.2$  \\ 
\hline
 $W^+ Z$ &    $15.8^{\footnotesize +4.1\%}_{\footnotesize -5.1\%}$    &     $28.3^{\footnotesize +4.9\%}_{\footnotesize -3.9\%}$      &  $31.3^{\footnotesize +2.3\%}_{\footnotesize -2.0\%}$  &    $1.8$   &   $1.1$  \\ 
\hline
 $W^- Z$ &    $9.7^{\footnotesize +4.5\%}_{\footnotesize -5.5\%}$     &     $17.8^{\footnotesize +4.9\%}_{\footnotesize -4.0\%}$      &  $19.8^{\footnotesize +2.2\%}_{\footnotesize -2.0\%}$  &    $1.8$   &   $1.1$  \\ 
\hline
\end{tabular}
\caption{Total cross section in pb at $13\tev$ LHC for on-shell production, with 
the center scale $\mu_R = \mu_F = (M_V + M_{V'})/2$.
Statistical errors are beyond the digits shown. 
The errors here are scale uncertainties due to variation of factor 2 around the center scale. 
Results taken from Refs.~\cite{Cascioli:2014yka,Gehrmann:2014fva,Grazzini:2016swo}. 
NLO EW corrections are negligible for total cross section, see \sect{sec:NLO_EW}.}
\label{tab:total_xsection}
\end{center}
\end{table}

From this table, we observe that NLO QCD corrections to the total cross section are large, about $50\%$ for 
the $ZZ$ and $WW$ processes and $80\%$ for $W^\pm Z$. We can separate the NLO correction into two parts: 
(i) the $q\bar{q}$ part includes the virtual and real-gluon-radiated corrections and
(ii) the $qg$ part includes the quark-gluon induced corrections. They are individually UV and IR finite. 
The calculation of \bib{Baglio:2013toa} reveals that the $q\bar{q}$ part is completely dominant for $ZZ$ and $WW$. 
The $qg$ correction is very small (less than $10\%$ compared to the Born) because of a cancellation between different 
regions of phase space. This can be seen in \fig{fig:dist_pT_NLO_QCD_ZZ_WW}, where the correction is positive 
at large $p_T$ and negative at low energies. The case of $W^\pm Z$ is different. 
The $q\bar{q}$ correction is still larger, but no longer dominant. 
The $qg$ correction is large because the cancellation between 
low and high energy regions is less severe, as can be seen in \fig{fig:diag_NLO_QCD_with_pT_WZ} (right). 
And since both corrections are positive, the total correction is larger than in the $ZZ$ and $WW$ cases.

We now turn to the NNLO QCD corrections. Compared to the NLO QCD cross section, 
the correction is about $20\%$ for $ZZ$ and $WW$ and $10\%$ for $W^\pm Z$, as shown in \tab{tab:total_xsection}. 
The correction for the electrically neutral final states is larger, partly due to the 
loop-induced $gg$ mechanism, which introduces a positive correction of about $8(3)\%$ for $ZZ(WW)$, compared to 
the NLO result at $14\tev$ LHC \cite{Baglio:2013toa}.

\begin{figure}[htb]
\begin{center}
\begin{minipage}[c]{6.5cm}
\epsfig{file=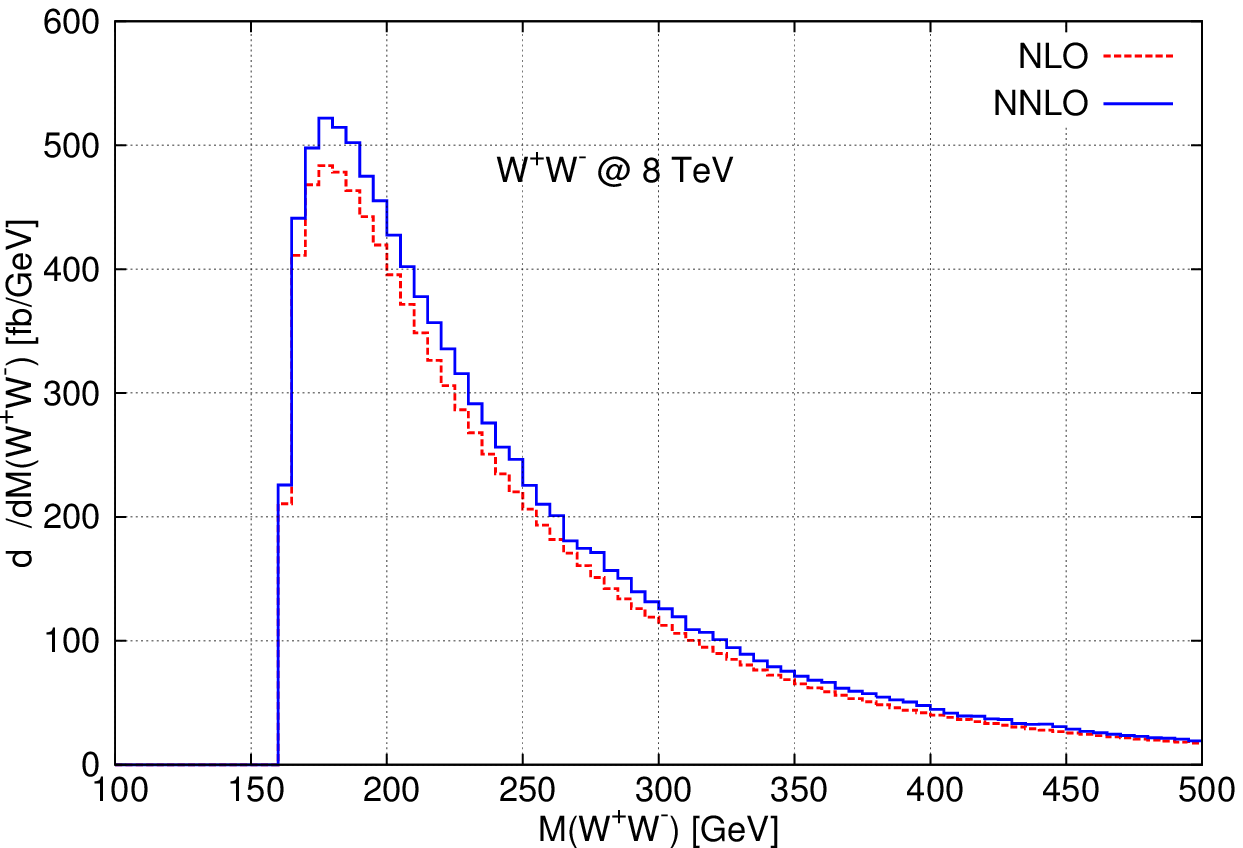,width=6.1cm}
\end{minipage}
\begin{minipage}[c]{6.0cm}
\epsfig{file=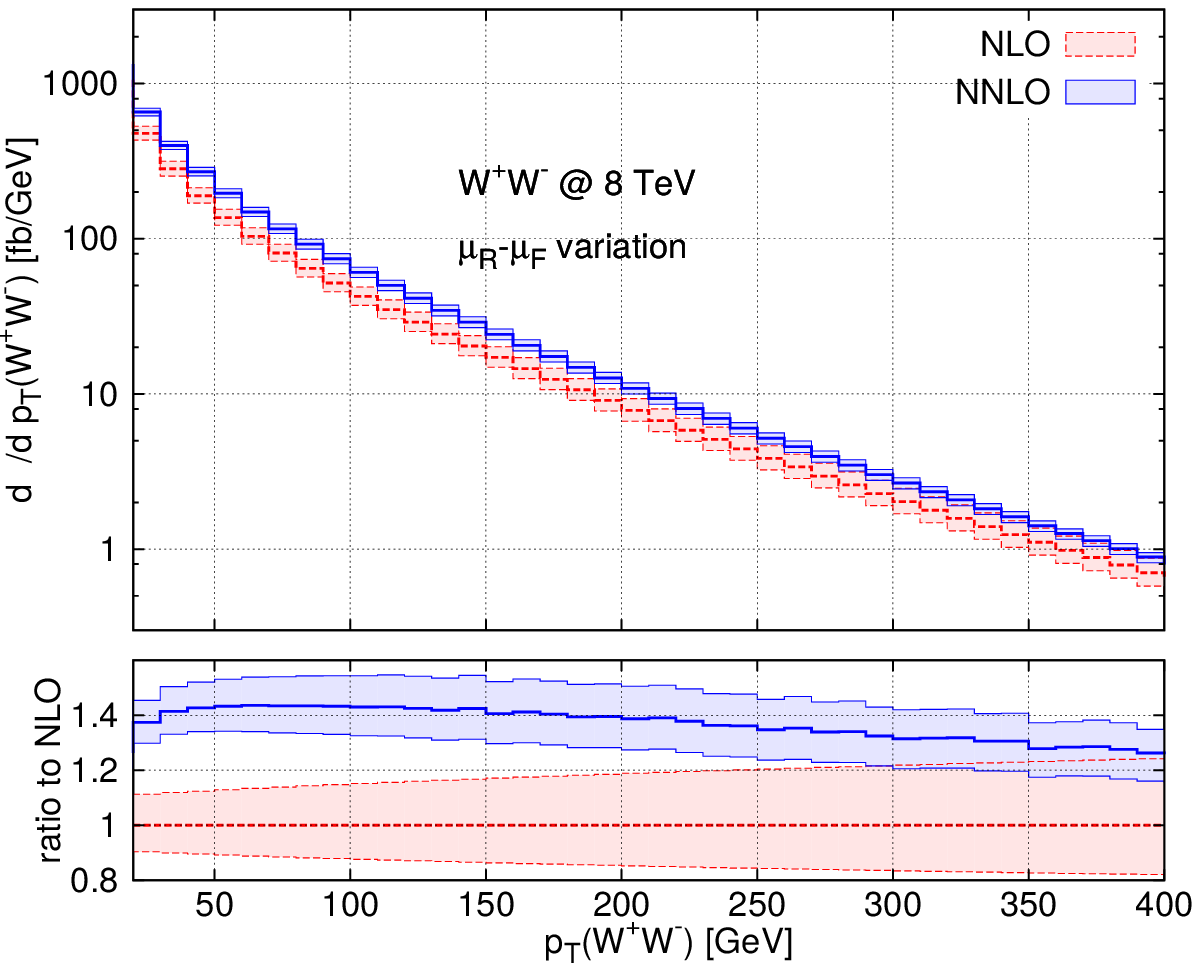,width=6.0cm}
\end{minipage}
\caption{Invariant mass (left) and transverse momentum (right) distributions 
of the $WW$ system for on-shell $WW$ production. 
Plots taken from \bib{Grazzini:2015wpa}.}
\label{fig:dist_NNLO_QCD_WW}
\end{center}
\end{figure}
We move on to distributions. In \fig{fig:dist_NNLO_QCD_WW}, 
invariant mass and transverse momentum distributions 
of the $WW$ system, for on-shell $WW$ production at $8\tev$ LHC, are shown. 
The NNLO correction is large at 
low energies (reaching $40\%$ at $p_T\approx 50\gev$), where both gauge bosons are soft. 
This may be due to the $gg \to \bar{q}q'VV'$ mechanism above mentioned. 

\section{Uncertainties and precision}
We first discuss theoretical uncertainties on the total cross section.
The common way to {\it estimate} this is combining the scale uncertainty with the 
PDF+$\alpha_s$ uncertainty (you may call it a sophisticated guess). The scale uncertainties are shown in 
\tab{tab:total_xsection}. The PDF+$\alpha_s$ uncertainties 
are about $4\%$ for all three processes according to \bib{Baglio:2013toa}. 
ATLAS used $2$ or $3\%$ \cite{Aad:2016wpd,Aaboud:2016yus} and CMS about $1.5\%$ \cite{Khachatryan:2016tgp,Khachatryan:2016txa}. 
To be conservative, we choose to use $4\%$.
Adding them linearly gives about $7\%$ uncertainty at NNLO QCD. 
This gives us an idea of the highest precision we can achieve. 
Compared to LEP2 \cite{Schael:2013ita}, it is better for $ZZ$ but worse for $WW$. 

The theoretical uncertainty should be taken with a pinch of salt, as can be seen from 
\tab{tab:total_xsection}. Referencing to the NNLO result, we see that both the 
LO and NLO scale uncertainties are too small. So, it is better to be a bit more conservative.

Can we compare theory and experiment calculations? 
For this we have to understand the experiment calculation. 
After the reconstruction of charged leptons and jets, passing a set 
of kinematic cuts which define the fiducial phase space, a signal cross section is 
calculated. This is called fiducial cross section, $\sigma_\text{fid}^\text{ex}$. 
We have to use the same cuts for leptons and jets in the theory calculation to get 
$\sigma_\text{fid}^\text{th}$. The two observables are not exactly the same, because a jet in 
experiment is reconstructed from calorimeter deposits while it is a combination of partons in theory. 
Similar things can be said for leptons. 
It can be helpful to think of the experiment measurement as an all-order calculation. 
We therefore should not expect a perfect agreement. 
At the moment, the agreements among ATLAS, CMS and theory are 
within two standard deviations \cite{Khachatryan:2015sga,Aad:2015zqe,Aad:2016wpd,Aaboud:2016yus,Khachatryan:2016tgp,Khachatryan:2016txa}. 
However, the precision is not very high, larger than $10\%$ (which is not 
a surprise given the above large theory uncertainty). From the experimental side, statistical errors 
can be reduced with high luminosity, but systematic errors are difficult to reduce. 

The difficulty with fiducial cross section comparisons is that the fiducial phase space is often changing with time (depending on 
machine energy, choice of analysis, ...) and varies with experiments. Moreover, what should we do if we want to combine results 
of different experiments? It is therefore useful to calculate also the total cross section based on a common definition of 
the total phase space. For this we have to extrapolate from the fiducial to the total phase space. 
This extrapolation factor is calculated using Monte-Carlo programs, assuming the SM. 
This is not a good thing to do if we want to search for new physics, please keep this in mind.

\section{NLO EW corrections}
\label{sec:NLO_EW}
Since $\alpha \approx \alpha_s^2$, we naively expect that NLO EW corrections are 
about the size of NNLO QCD ones. 
For the total cross section, we have seen that 
NNLO QCD brings $10\%$ for $ZZ$ or $WW$ and $20\%$ for $W^\pm Z$. 
NLO EW correction is about $-3\%$ for $ZZ$, $+1.5\%$ for $WW$ (mainly due to 
the LO $\gamma\gamma$ contribution counted as a NLO EW correction here), 
and is negligible for $WZ$ channels. To understand why NLO EW corrections to the 
total cross section are so small, we have to look at distributions.  

\begin{figure}[htb]
\epsfig{file=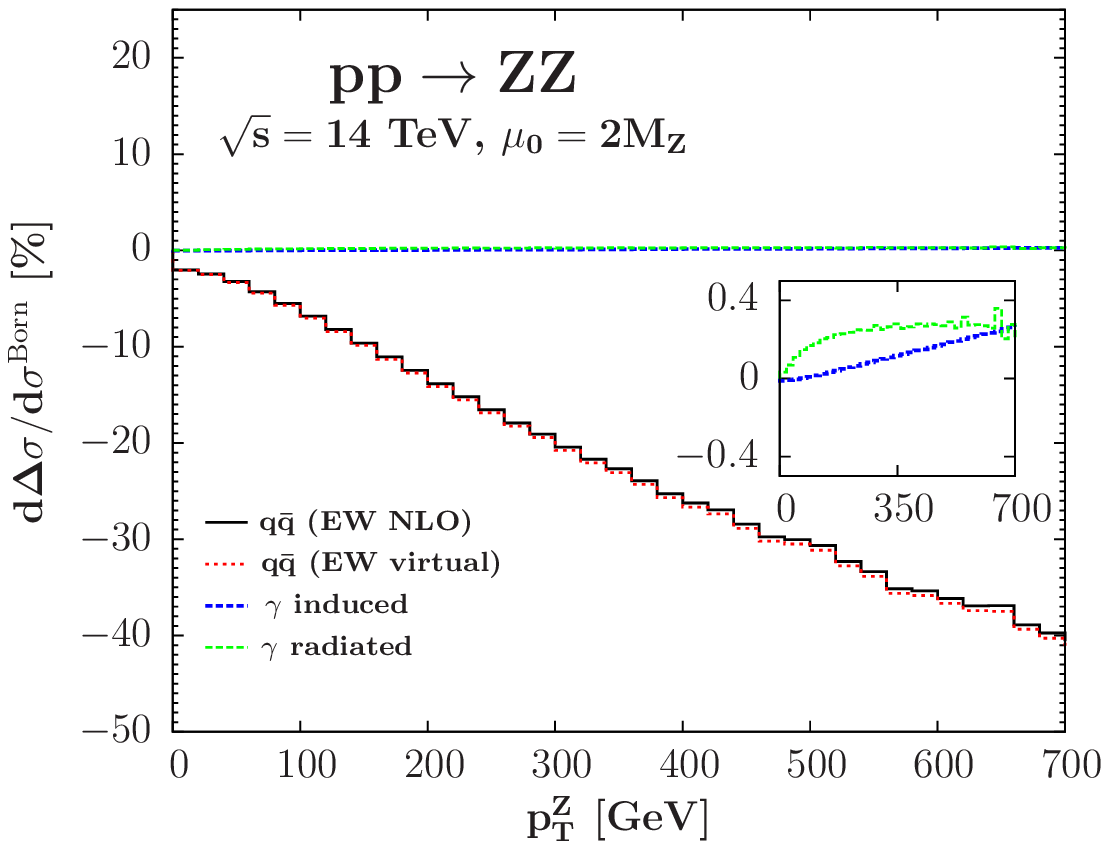,width=4.9cm}
\epsfig{file=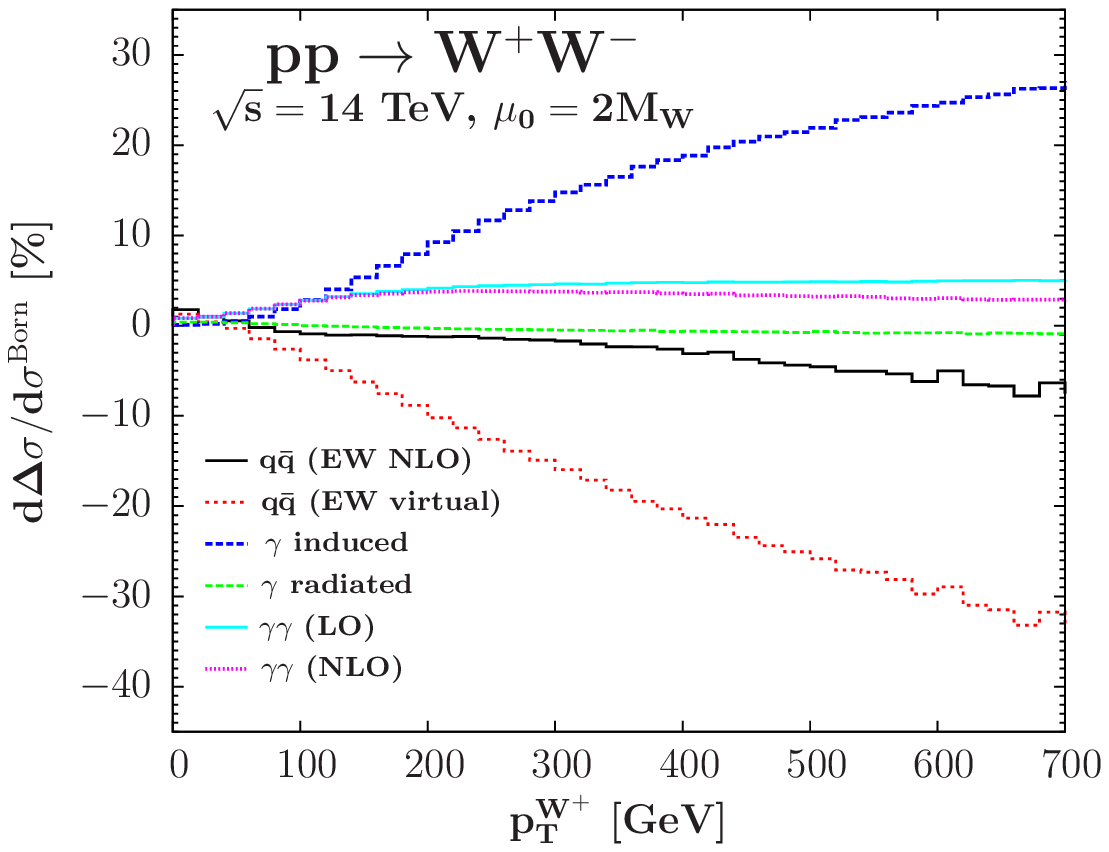,width=4.9cm}
\epsfig{file=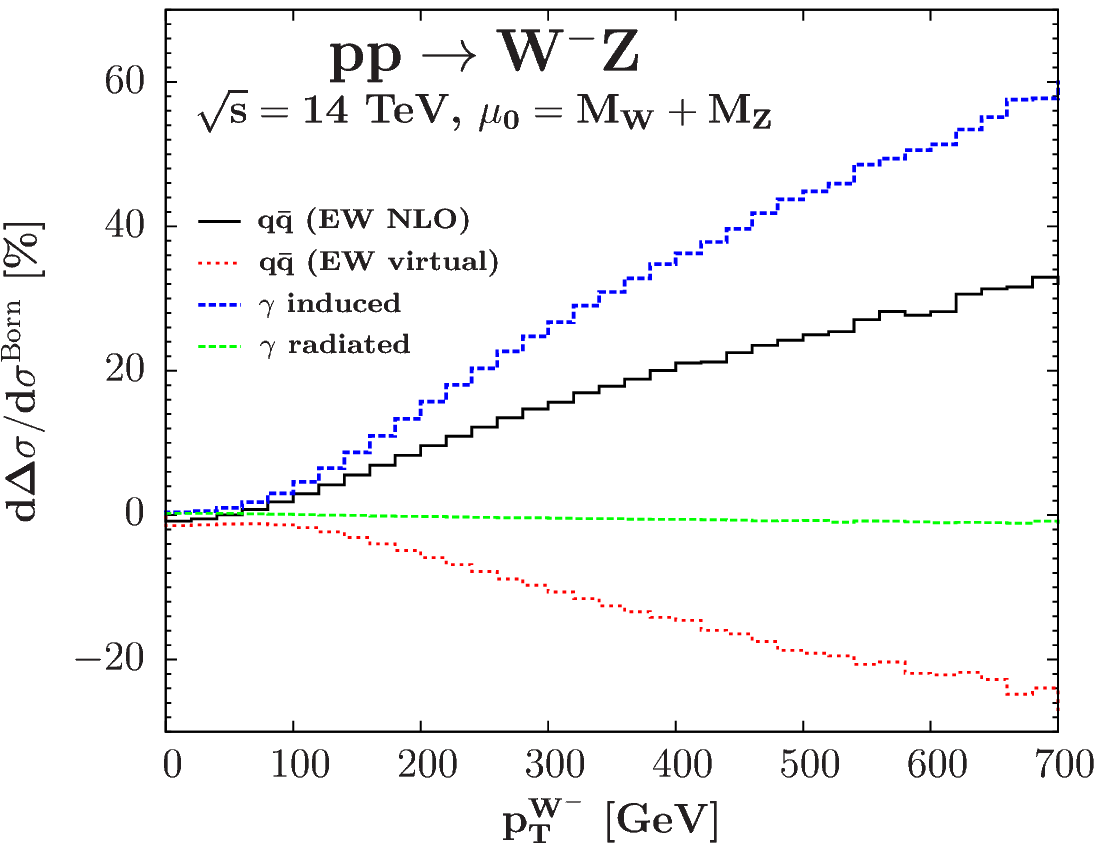,width=4.95cm}
\caption{Transverse momentum distribution for $ZZ$ (left), $WW$ (middle), and $W^- Z$ (right).
Plots taken from \bib{Baglio:2013toa}.}
\label{fig:dist_pT_NLO_EW_VV}
\end{figure}
\begin{figure}[ht]
\begin{center}
\epsfig{file=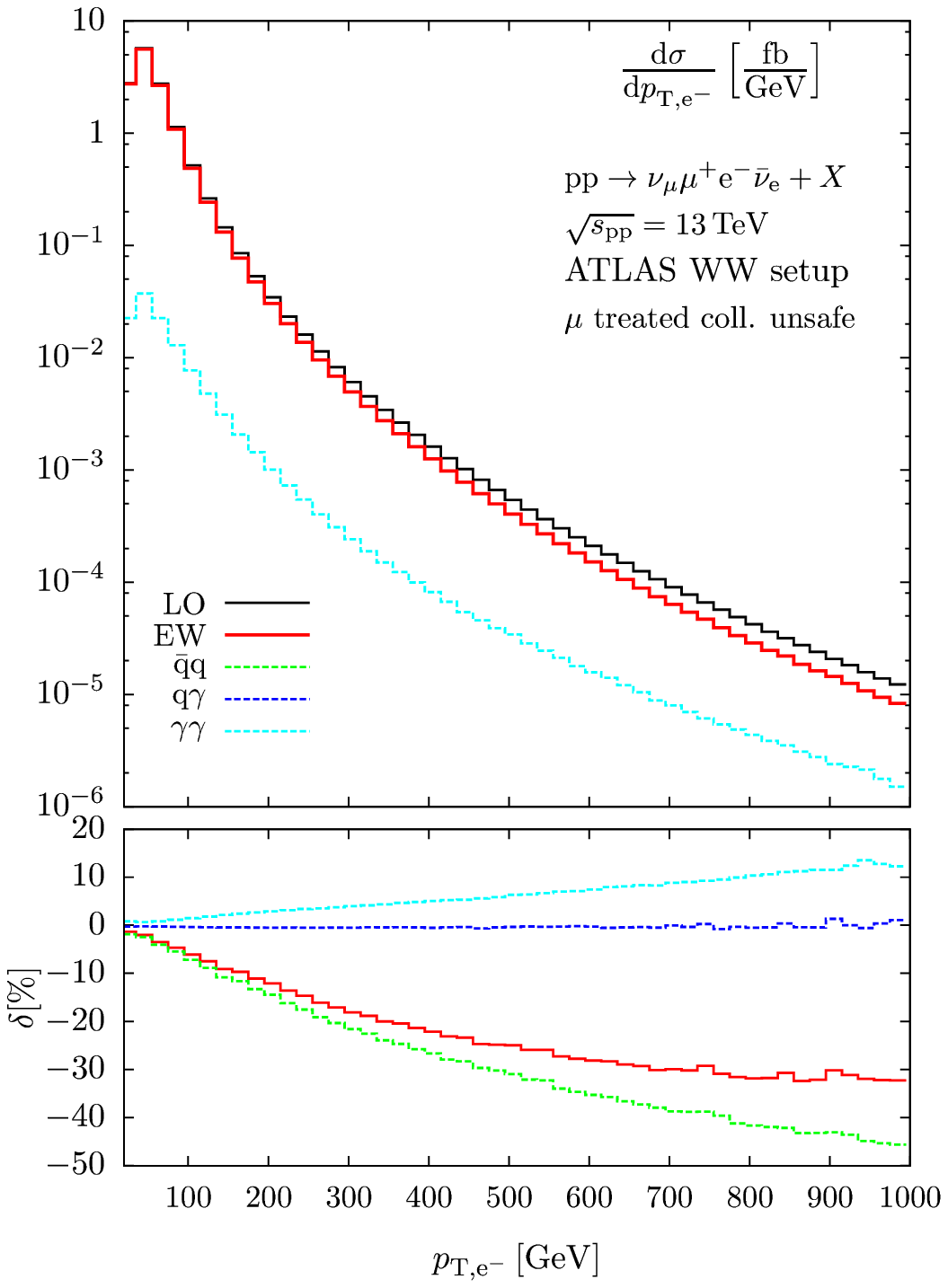,width=6.0cm}
\epsfig{file=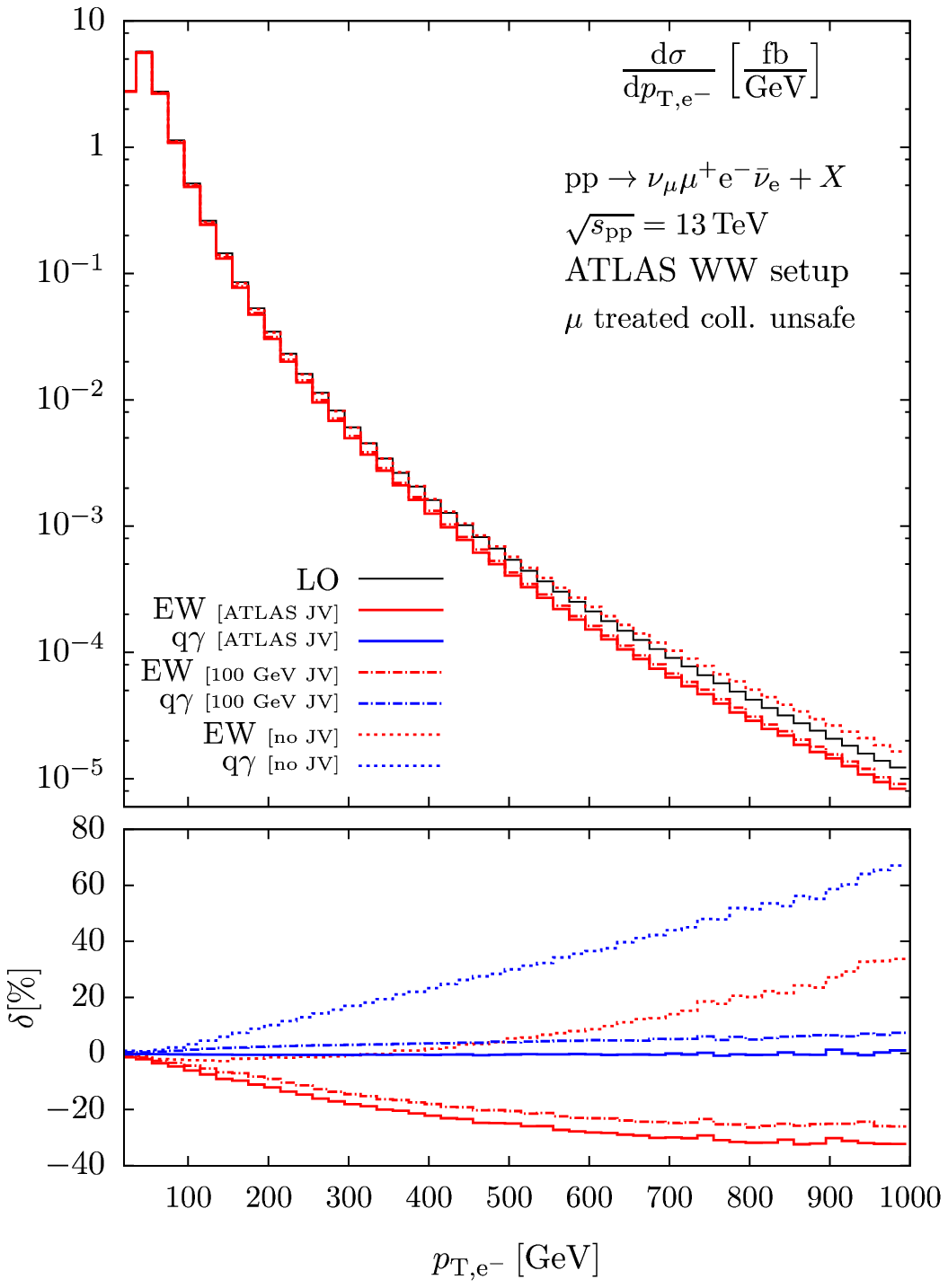,width=6.0cm}
\caption{Transverse momentum distributions of the electron with jet veto of $p_\text{veto} = 25\gev$ (left) 
and with jet-veto (JV) dependence (right) for $\nu_\mu \mu^+ e^- \bar{\nu}_e$ production at $13\tev$ LHC. 
Note that the label [ATLAS JV] means $p_\text{veto} = 25\gev$ as in the left plot. 
The relative size of various corrections are shown in the lower panels. 
Plots taken from \bib{Biedermann:2016guo} where cuts and PDFs are specified.}
\label{fig:dist_NLO_EW_WW4l}
\end{center}
\end{figure}

In \fig{fig:dist_pT_NLO_EW_VV} some representative $p_T$ distributions for $ZZ$ (left), $WW$ (middle), and $W^- Z$ (right) are displayed. Similarly to NLO QCD corrections, we split the NLO EW correction into 
different contributions: virtual, quark-photon induced, and photon-radiated corrections. 
The photon-photon contribution at LO and NLO to $WW$ production are also plotted. 
We observe that
the effect of the virtual Sudakov logarithms $\alpha
\log^2[(p_T^{Z})^2/M_W^2]$ is clearly visible in
the $ZZ$ panel, reaching $-40\%$ correction at $p_T^{Z}=700\gev$. This effect is 
also important for $WW$ and $WZ$. However, it is largely cancelled by the quark-photon induced contribution. 
We notice that the $q\gamma$ contribution is largest for $W^- Z$, reaching $+60\%$ at $700\gev$, about 
$+25\%$ for $WW$ and negligible for $ZZ$. This can be explained using the same argument presented above for the quark-gluon induced contribution. It is very small for $ZZ$ because the photon PDF is suppressed. 
It is large for $WW$ and $WZ$ because there occurs a new mechanism related to the EW $\gamma \to W^+ W^-$ splitting, which introduces a hard process with $t$-channel $W$ exchange. The amplitude of this new hard process is large, leading to large corrections. An analytical explanation using leading logarithmic approximation is given in \bib{Baglio:2013toa}. The cancellation of the virtual and quark-photon induced 
contributions in the case of $WW$ and $WZ$ is really striking. We will aslo see this when leptonic decays are included (see the right plot of \fig{fig:dist_NLO_EW_WW4l}). It is important to keep this cancellation in mind. The photon contribution is more important than we thought.   

\begin{figure}[htb]
\begin{center}
\epsfig{file=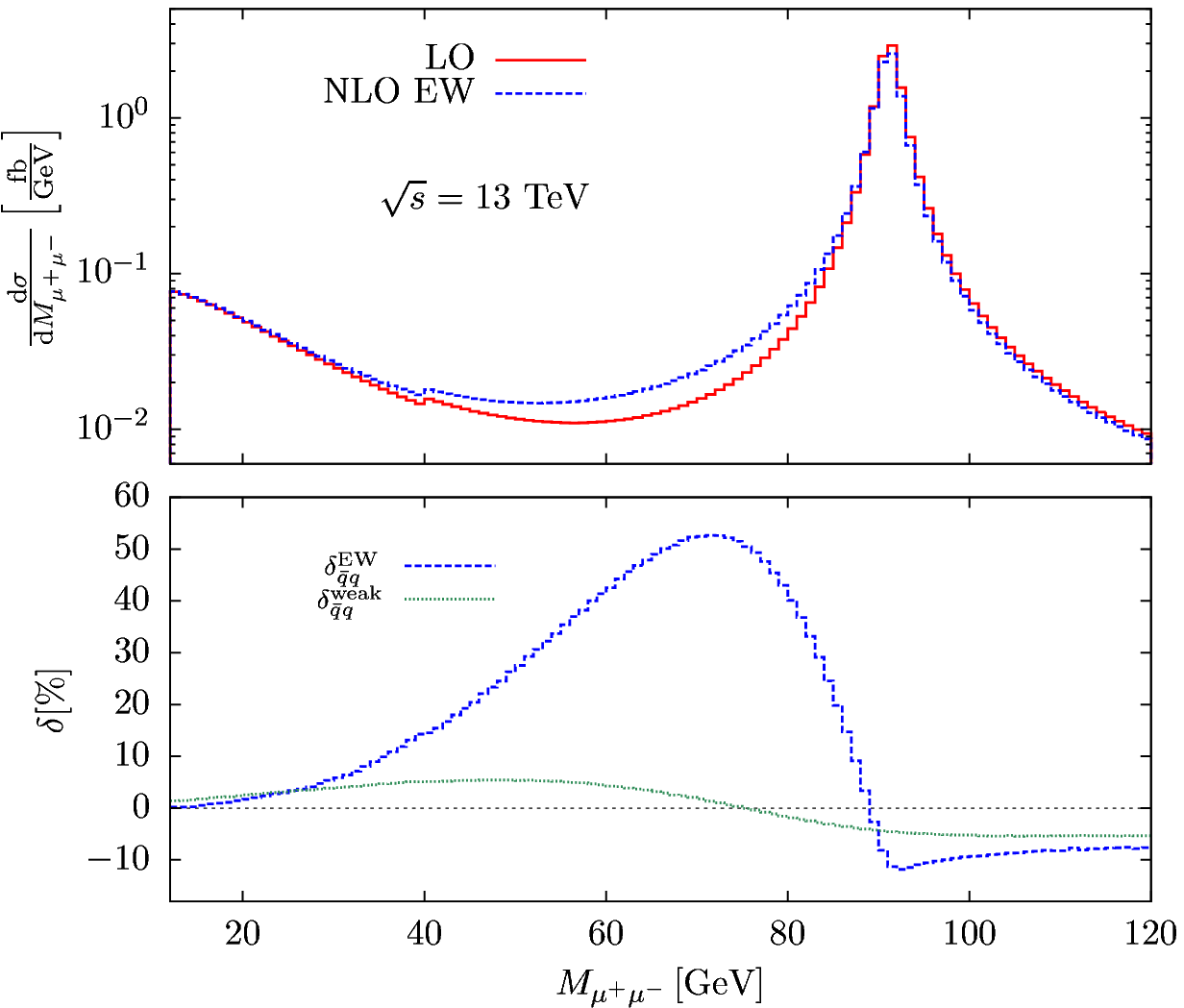,width=6.5cm}
\epsfig{file=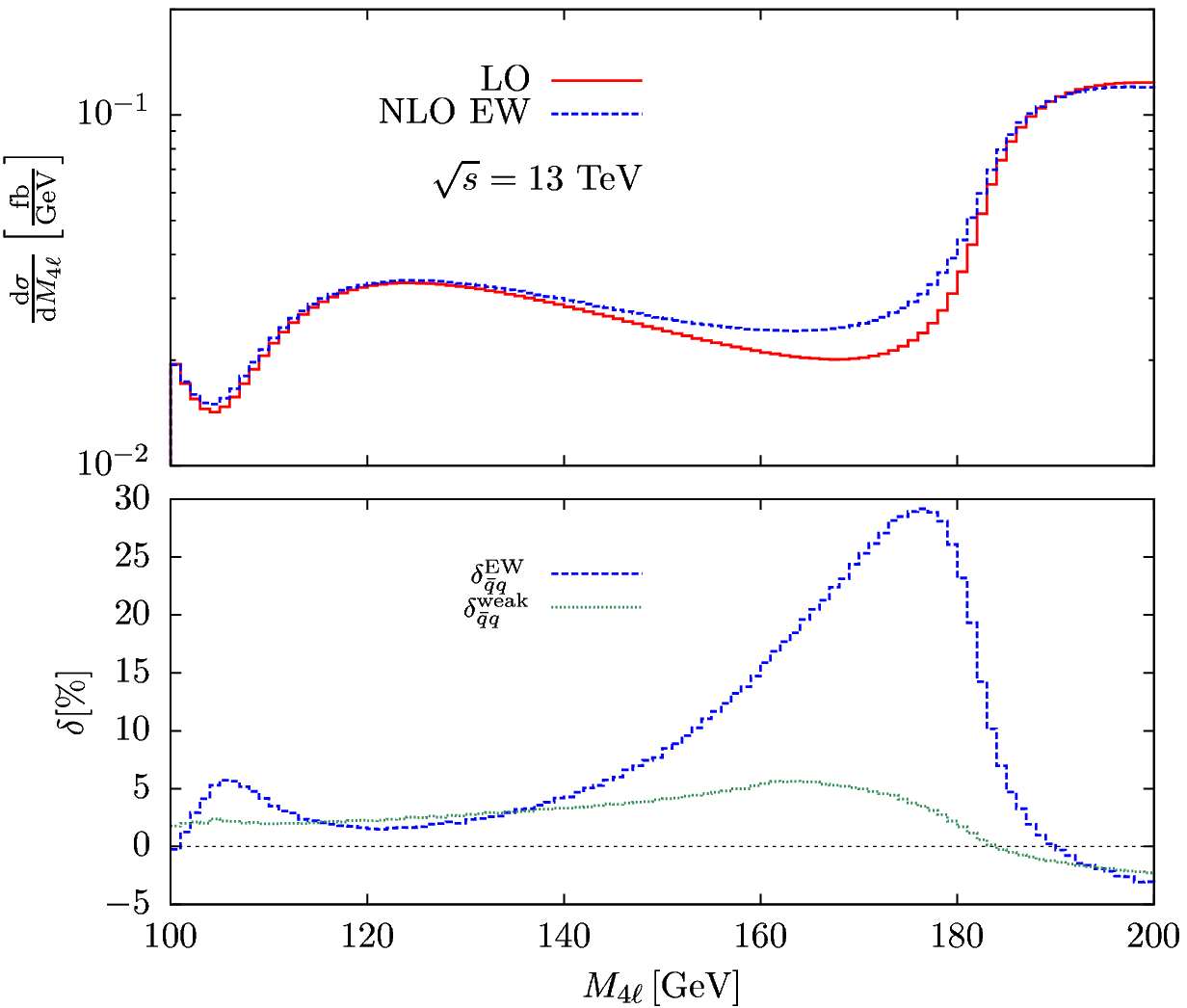,width=6.5cm}
\caption{Invariant mass distributions of $\mu^+ \mu^-$ (left) 
and of $\mu^+ \mu^- e^+ e^-$ (right) system 
for $\mu^+ \mu^- e^+ e^-$ production at $13\tev$ LHC. 
Cuts are used to require $M_{\mu\mu}$ and $M_{ee}$ to be close to the $Z$ mass, 
with $M_{4l} > 100\gev$ in addition. 
The relative EW and purely weak corrections are shown in the lower panels.    
Plots taken from \bib{Biedermann:2016yvs} where other cuts and PDFs are specified.}
\label{fig:dist_NLO_EW_ZZ4l}
\end{center}
\end{figure}
We now look at \fig{fig:dist_NLO_EW_ZZ4l}, where the process $pp \to \mu^+ \mu^- e^+ e^- + X$ 
is considered with full NLO EW corrections \cite{Biedermann:2016yvs}. The charged leptons 
originate from either the $Z$ boson or the photon. 
However, cuts are used to require $M_{\mu\mu}$ and $M_{ee}$ to be close to the $Z$ mass, 
with $M_{4l} > 100\gev$ in addition. The latter means that dominant contributions come from kinematic configurations where 
at least one $Z$ can be on-shell. 
The figure shows the $M_{\mu\mu}$ (left) and $M_{4l}$ distributions, together 
with the full NLO EW corrections and the genuine 
weak corrections (where photonic corrections are excluded) in the lower panels. 
We see that the photonic corrections can be large, reaching maximally $+50\%$ in $M_{\mu\mu}$ distribution 
and about $+25\%$ in $M_{4l}$. These large QED corrections occur just before the $M_Z$ and $2M_Z$ thresholds. 
This is due to final-state-radiation effects, which introduce new on-shell poles at lower values 
of $M_{\mu\mu}$ or $M_{4l}$ in the phase space, as noted in \bib{Biedermann:2016yvs}. 
It is also interesting to note that both corrections change the sign near the thresholds. 
From a practical viewpoint, it is important to separate the purely weak corrections so that 
we can fetch it into a parton-shower Monte-Carlo program to better account for photonic corrections. 
 
Full NLO EW corrections to $W^+W^-$ production, with leptonic decays, 
full off-shell and spin-correlation effects taken into account, have been 
recently calculated in \bib{Biedermann:2016guo}. Their results for $p_{T,e}$ distribution 
are shown in \fig{fig:dist_NLO_EW_WW4l}. For the left plot, 
a jet veto of $p_\text{veto} = 25\gev$ is used. 
The right plot displays additionally results for no jet veto 
and for $p_\text{veto} = 100\gev$. 
First, consider the case of no jet veto. We can compare the right plot with the $p_T^{W^+}$ distribution discussed 
above at the on-shell $WW$ level. The two plots look similar, both showing a strong cancellation between the 
quark-photon induced contribution and the virtual correction, leading to a small total EW correction up to $600\gev$. 
If jet veto is used then the quark-photon induced contribution is suppressed, making the total NLO 
EW correction more negative. 
The tight veto of $25\gev$ kills this almost completely, while the $100\gev$ veto also 
reduces it drastically, from $40\%$ to below $10\%$ at about $700\gev$. 
The motivation to use jet veto is to reduce the huge $t\bar{t}$ background. 
However, the price to pay is that theoretical uncertainties increase (see e.g. \cite{Stewart:2011cf}).

\section{Final remarks}
Can we reach $1\%$ precision for massive gauge boson pair production at the LHC? 
At the moment it looks impossible, the NNLO QCD uncertainties are still too large. 
We would need then next-to-NNLO QCD calculations and reduce PDF$+\alpha_s$ errors at the same time. 

In any case, matching and merging between
fixed-order calculations and parton shower at NNLO QCD level are mandatory. 
This is yet to be done.

\bigskip
{\noindent\bf Acknowledgments:}
I am grateful to the organizers for the invitation. 
I thank Rencontres du Vietnam for financial support and, in particular, 
Tran Thanh Van for being an inspiration. Some Feynman diagrams are drawn using Jaxodraw \cite{Binosi:2003yf}.


\begin{thebibliography}{99}

\bibitem{Aad:2012tfa}
ATLAS, G.~Aad {\em et~al.},
\newblock Phys. Lett. {\bf B716}, 1 (2012), 1207.7214.

\bibitem{Chatrchyan:2012xdj}
CMS, S.~Chatrchyan {\em et~al.},
\newblock Phys. Lett. {\bf B716}, 30 (2012), 1207.7235.

\bibitem{LEP-2}
{The ALEPH, DELPHI, L3, OPAL Collaborations, the LEP Electroweak Working
  Group},
\newblock Phys. Rept. {\bf 532}, 119 (2013), 1302.3415.

\bibitem{Mele:1990bq}
B.~Mele, P.~Nason, and G.~Ridolfi,
\newblock Nucl. Phys. {\bf B357}, 409 (1991).

\bibitem{Ohnemus:1990za}
J.~Ohnemus and J.~F. Owens,
\newblock Phys. Rev. {\bf D43}, 3626 (1991).

\bibitem{Ohnemus:1991kk}
J.~Ohnemus,
\newblock Phys. Rev. {\bf D44}, 1403 (1991).

\bibitem{WZ-QCD-NLO-1}
J.~Ohnemus,
\newblock Phys.Rev. {\bf D44}, 3477 (1991).

\bibitem{WZ-QCD-NLO-2}
S.~Frixione, P.~Nason, and G.~Ridolfi,
\newblock Nucl.Phys. {\bf B383}, 3 (1992).

\bibitem{WW-QCD-NLO}
S.~Frixione,
\newblock Nucl.Phys. {\bf B410}, 280 (1993).

\bibitem{Cascioli:2014yka}
F.~Cascioli {\em et~al.},
\newblock Phys. Lett. {\bf B735}, 311 (2014), 1405.2219.

\bibitem{Gehrmann:2014fva}
T.~Gehrmann {\em et~al.},
\newblock Phys. Rev. Lett. {\bf 113}, 212001 (2014), 1408.5243.

\bibitem{Grazzini:2016ctr}
M.~Grazzini, S.~Kallweit, S.~Pozzorini, D.~Rathlev, and M.~Wiesemann,
\newblock JHEP {\bf 08}, 140 (2016), 1605.02716.

\bibitem{Grazzini:2016swo}
M.~Grazzini, S.~Kallweit, D.~Rathlev, and M.~Wiesemann,
\newblock Phys. Lett. {\bf B761}, 179 (2016), 1604.08576.

\bibitem{WW-EW-Kuhn}
A.~Bierweiler, T.~Kasprzik, H.~{K\"uhn}, and S.~Uccirati,
\newblock JHEP {\bf 1211}, 093 (2012), arXiv:1208.3147.

\bibitem{Bierweiler:2013dja}
A.~Bierweiler, T.~Kasprzik, and J.~H. Kühn,
\newblock JHEP {\bf 12}, 071 (2013), 1305.5402.

\bibitem{Baglio:2013toa}
J.~Baglio, L.~D. Ninh, and M.~M. Weber,
\newblock Phys. Rev. {\bf D88}, 113005 (2013), 1307.4331,
\newblock [Erratum: Phys. Rev.D94,no.9,099902(2016)].

\bibitem{Biedermann:2016yvs}
B.~Biedermann, A.~Denner, S.~Dittmaier, L.~Hofer, and B.~Jäger,
\newblock Phys. Rev. Lett. {\bf 116}, 161803 (2016), 1601.07787.

\bibitem{Biedermann:2016guo}
B.~Biedermann {\em et~al.},
\newblock JHEP {\bf 06}, 065 (2016), 1605.03419.

\bibitem{Olive:2016xmw}
Particle Data Group, C.~Patrignani {\em et~al.},
\newblock Chin. Phys. {\bf C40}, 100001 (2016).

\bibitem{Aaboud:2016uuk}
ATLAS, M.~Aaboud {\em et~al.},
\newblock (2016), 1609.05122.

\bibitem{CMS:2016djf}
CMS, C.~Collaboration,
\newblock (2016).

\bibitem{Aad:2015rpa}
ATLAS, G.~Aad {\em et~al.},
\newblock Eur. Phys. J. {\bf C76}, 154 (2016), 1510.05821.

\bibitem{QCD-NLO-exclusive}
J.~Ohnemus,
\newblock Phys.Rev. {\bf D50}, 1931 (1994), hep-ph/9403331.

\bibitem{Grazzini:2015wpa}
M.~Grazzini, S.~Kallweit, D.~Rathlev, and M.~Wiesemann,
\newblock JHEP {\bf 08}, 154 (2015), 1507.02565.

\bibitem{Aad:2016wpd}
ATLAS, G.~Aad {\em et~al.},
\newblock JHEP {\bf 09}, 029 (2016), 1603.01702.

\bibitem{Aaboud:2016yus}
ATLAS, M.~Aaboud {\em et~al.},
\newblock Phys. Lett. {\bf B762}, 1 (2016), 1606.04017.

\bibitem{Khachatryan:2016tgp}
CMS, V.~Khachatryan {\em et~al.},
\newblock Phys. Lett. {\bf B} (2016), 1607.06943.

\bibitem{Khachatryan:2016txa}
CMS, V.~Khachatryan {\em et~al.},
\newblock Phys. Lett. {\bf B763}, 280 (2016), 1607.08834.

\bibitem{Schael:2013ita}
DELPHI, OPAL, LEP Electroweak, ALEPH, L3, S.~Schael {\em et~al.},
\newblock Phys. Rept. {\bf 532}, 119 (2013), 1302.3415.

\bibitem{Khachatryan:2015sga}
CMS, V.~Khachatryan {\em et~al.},
\newblock Eur. Phys. J. {\bf C76}, 401 (2016), 1507.03268.

\bibitem{Aad:2015zqe}
ATLAS, G.~Aad {\em et~al.},
\newblock Phys. Rev. Lett. {\bf 116}, 101801 (2016), 1512.05314.

\bibitem{Stewart:2011cf}
I.~W. Stewart and F.~J. Tackmann,
\newblock Phys.Rev. {\bf D85}, 034011 (2012), arXiv:1107.2117.

\bibitem{Binosi:2003yf}
D.~Binosi and L.~Theussl,
\newblock Comput. Phys. Commun. {\bf 161}, 76 (2004), hep-ph/0309015.

\end{thebibliography}
\end{document}